\documentclass[preprint]{aastex631}

\begin{document}

\title{Redshifting the Study of Cold Brown Dwarfs and Exoplanets:\\ the {\bf Mid-Infrared} Wavelength Region as an Indicator of Surface Gravity and Mass}

\correspondingauthor{Sandy Leggett}
\email{sandy.leggett@noirlab.edu}

\author[0000-0002-3681-2989]{S. K. Leggett}
\affiliation{Gemini Observatory/NSF's NOIRLab, 670 N. A'ohoku Place, Hilo, HI 96720, USA}

\author[0000-0001-6172-3403]{Pascal Tremblin}
\affiliation{Universite Paris-Saclay, UVSQ, CNRS, CEA, Maison de la Simulation, 91191, Gif-sur-Yvette, France
}

%% Mark off the abstract in the ``abstract'' environment. 

\begin{abstract}

{\it JWST} is opening many avenues for  exploration. For cold brown dwarfs and exoplanets, {\it JWST} has opened the door to the mid-infrared wavelength region, where such objects emit significant energy. For the first time, astronomers have access to mid-infrared spectroscopy for objects colder than 600~K. The first spectra appear to validate the model suite known as ATMO 2020++: atmospheres which include disequilibrium chemistry and have a non-adiabatic pressure-temperature relationship. 
Preliminary fits to {\it JWST} spectroscopy of Y dwarfs show that the slope of the energy distribution from $\lambda \approx 4.5~\mu$m to $\lambda \approx 10~\mu$m is very sensitive to gravity. We explore this phenomenon using PH$_3$-free ATMO 2020++ models and updated {\it WISE} W2 $-$ W3 colors.  We find that 
an absolute $4.5~\mu$m flux measurement constrains temperature, and the ratio of the $4.5~\mu$m flux to the 10 -- 15$~\mu$m flux is sensitive to gravity and less sensitive to metallicity.  We identify 10 T dwarfs with red W2 $-$ W3 colors which are likely to be very low gravity, young, few-Jupiter-mass objects; one of these is the previously known  COCONUTS-2b. The unusual Y dwarf WISEPA J182831.08+265037.8 is blue in W2 $-$ W3 and we find that the 4 to 18$~\mu$m {\it JWST} spectrum is well reproduced if the system is a pair of high gravity 400~K dwarfs. 
Recently published {\it JWST} colors and luminosity-based effective temperatures for late-T and Y dwarfs further corroborate the ATMO 2020++ models, demonstrating the potential for significant improvement in our understanding of 
cold very low-mass bodies in the solar neighborhood.

\end{abstract}

%% Keywords should appear after the \end{abstract} command. 
%% The AAS Journals now uses Unified Astronomy Thesaurus concepts:
%% https://astrothesaurus.org
%% You will be asked to selected these concepts during the submission process
%% but this old "keyword" functionality is maintained in case authors want
%% to include these concepts in their preprints.

\keywords{Brown dwarfs --- Exoplanet astronomy --- Fundamental parameters of stars --- Infrared photometry}

%\bigskip
%\clearpage
\section{Introduction} 

Wien's law of blackbody radiation states that the peak of the energy distribution will occur at wavelengths inversely proportional to temperature. Although stars and brown dwarfs (objects with a mass too low for stable hydrogen fusion e.g. \citet{Burrows_1993, Saumon_2008, Phillips_2020}) are not blackbodies, an increasing amount of energy is emitted at longer wavelengths as their effective temperature ($T_{\rm eff}$) decreases.  The known T- and Y-type brown dwarfs have $T_{\rm eff}$s approximately 1200 -- 500~K and 500 -- 250~K respectively \citep[e.g.][]{Kirkpatrick_2021a};  Figure 1 illustrates how the bulk of the energy emission shifts from the near-infrared to the mid-infrared as late-T dwarfs cool to Y dwarfs.

%\clearpage
\begin{figure}[!ht]
\vskip -0.2in
%\hskip -0.25in
%\plotone{lowResSED.pdf}
\includegraphics[angle=-90,width = 7.3 in]
{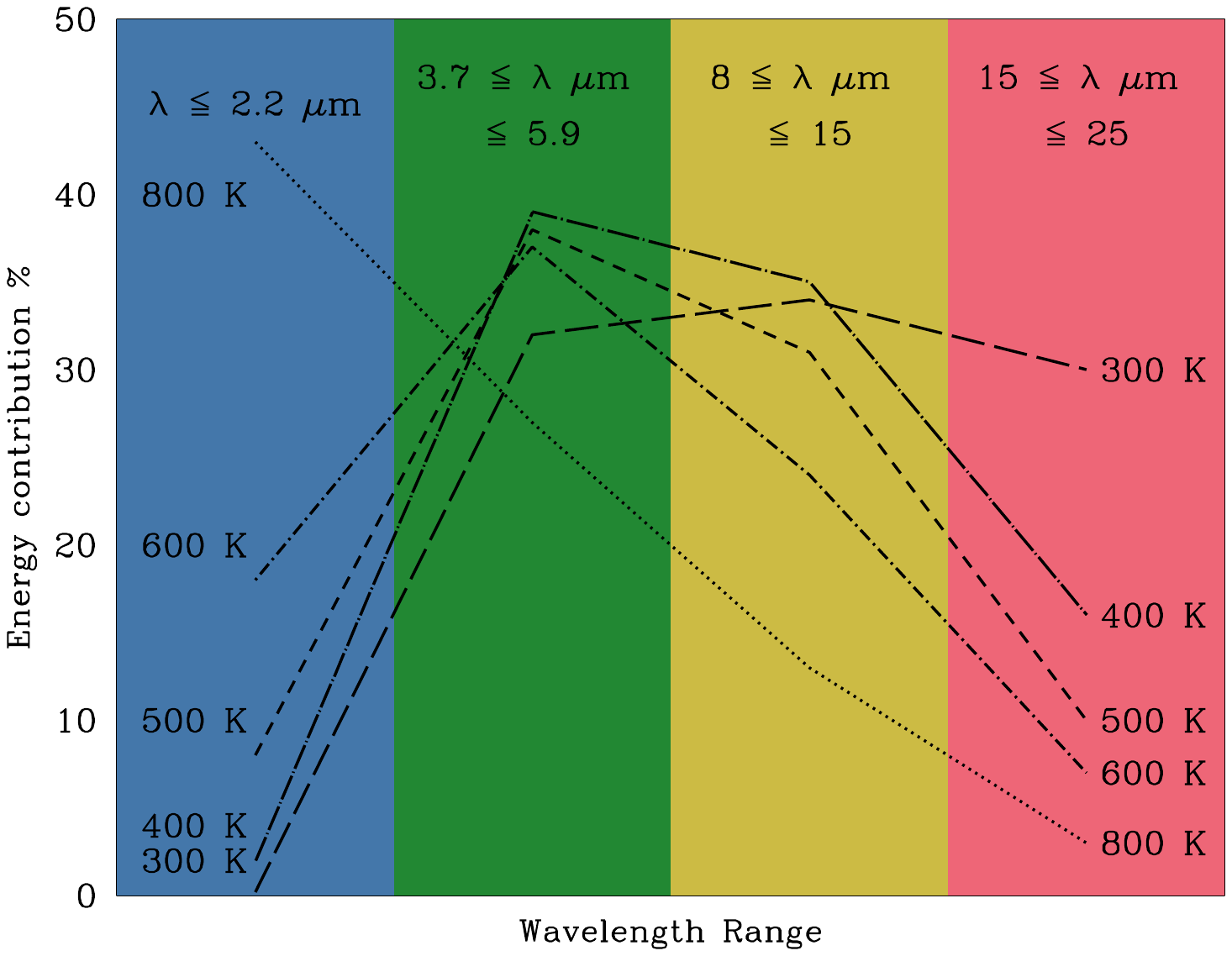}
%{W3_fnu.pdf}
\vskip -0.2in
\caption{Illustration of the energy distribution shift to longer wavelengths as $T_{\rm eff}$ decreases, for
late-T and Y dwarfs. The fractional energy contribution is estimated using ATMO 2020++ model atmospheres \citep{Phillips_2020, Leggett_2021} with solar metallicity and a surface gravity log $g = 4.5$. Note that approximately one-third of the total energy emerges at $3.7 \leq \lambda~\mu$m $\leq 5.9$, which corresponds to the {\it WISE} W2, {\it Spitzer} [4.5], or {\it JWST} F444W band.
For Y dwarfs, with $T_{\rm eff} \lesssim 500$~K, an additional third emerges at $8 \leq \lambda~\mu$m $\leq 15$, which corresponds to the {\it WISE} W3 band. 
}
\end{figure}

The {\em Wide-field Infrared Survey Explorer} 
\citep[{\em WISE},][]{Wright_2010} provided vital mid-infrared imaging data, enabling discovery and characterization of very cold brown dwarfs \citep[e.g.][]{Cushing_2011, Kirkpatrick_2011, Meisner_2020a, Leggett_2021}. A synthetic Y dwarf spectrum is shown in Figure 2 with filter profiles superimposed. The {\em WISE} W2 ($4.6~\mu$m) and W3 ($12~\mu$m) filters sample wavelength regions where significant flux is emitted (Figures 1 and 2).   
We note that the flux peak centered near $\lambda = 4.5~\mu$m lies in a window between absorption bands of CH$_4$ at $\lambda \sim 3~\mu$m and H$_2$O at $\lambda \sim 5~\mu$m \citep[e.g.][their Figure 7]{Morley_2014}, and is accessible by ground-based telescopes. The diagnostic power of this bandpass has been recognised for some time \citep[e.g.][]{Noll_1997,
Burrows_2003, Leggett_2007, Leggett_2010}.

%\clearpage
\begin{figure}[!ht]
\vskip -0.2in
%\hskip -0.25in
\includegraphics[angle=-90,width = 7.3 in]
{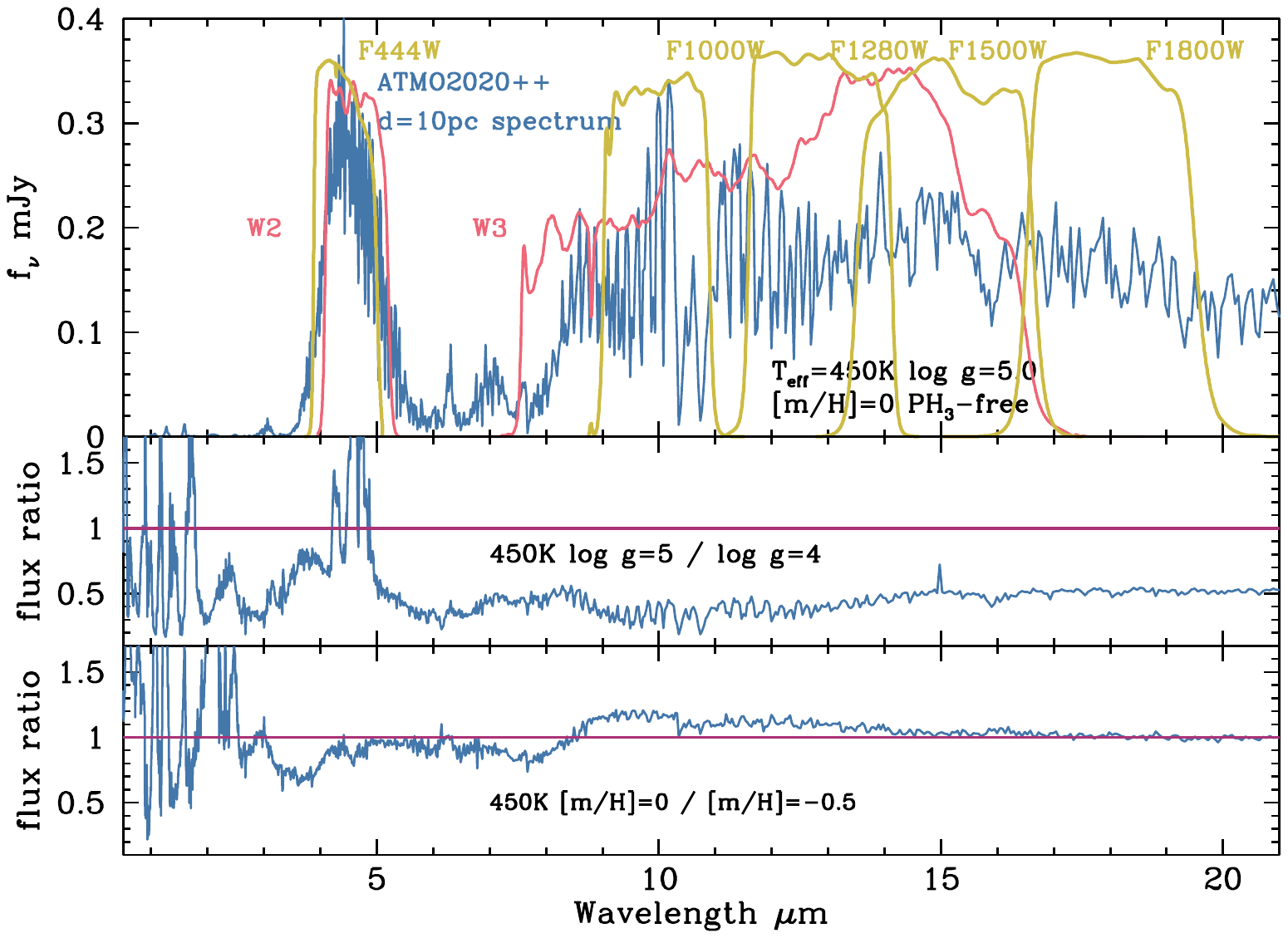}
%\vskip -0.4in
\caption{The top panel shows the synthetic spectrum for a Y dwarf at 10~pc with the atmospheric parameters given in the legend. {\it WISE} and {\it JWST} filter bandpasses are illustrated as red and yellow lines respectively. The middle and lower panels show the ratio of this spectrum divided by a spectrum where the surface gravity has been reduced, and by a spectrum where the metallicity is reduced, respectively.  The total energy emitted by the lower gravity  dwarf is greater because the radius is larger \citep{Marley_2021}, hence the average value in the middle panel is $< 1$. Note that all three models have similar flux levels averaged over the $\lambda \approx 4.5~\mu$m flux peak, for example through an imaging filter, although the strength of the absorption features at these wavelengths (primarily due to CO and CO$_2$) are sensitive to gravity. See Section 3 for further discussion.
}
\end{figure}

The {\em James Webb Space Telescope (JWST)} is now obtaining mid-infrared spectra of Y dwarfs \citep{Beiler_2023, Barrado_2023, Beiler_2024}. 
These are the first spectra of Y dwarfs at $\lambda > 5~\mu$m and they enable more rigorous testing of brown dwarf model atmospheres. The new data appear to validate 
the model suite known as ATMO 2020++: atmospheres which include disequilibrium chemistry and have a non-adiabatic pressure-temperature relationship
\citep{Tremblin_2019,Phillips_2020,Leggett_2021, Leggett_2023, Leggett_2024}. 
\citet{Leggett_2024} show preliminary fits of ATMO 2020++ synthetic spectra to low resolution {\it JWST} spectra. A striking feature is the dependency of the spectral shape on surface gravity $g$. In particular the relative heights of the $\lambda \approx 4.5~\mu$m and $\lambda \approx 10~\mu$m flux peaks are very different for  log $g = 4.0$ and log $g = 4.5$ at $T_{\rm eff} = 450$~K. This observation led to the exploration of the dependency of the {\it WISE} W2 and W3 colors on gravity, presented here. 

Section 2 describes the sample and presents updated W3 magnitudes. Section 3 presents our analysis, which includes identification of candidate low gravity (very low mass) T dwarfs, and exploration of the unusual Y dwarf WISEPA J182831.08+265037.8 as an example of a high gravity system. Section 4 applies the technique to a set of {\it JWST} colors, and shows that the observed colors and luminosities recently published by \citet{Beiler_2024}, for a sample of 22 late-T and Y dwarfs, further corroborates the ATMO 2020++ models. Section 5 gives our conclusions. The Appendix provides our photometric data compilation.

\bigskip
\section{Sample and Data}

For this study we started with the photometry compilation for late-T and Y dwarfs presented in Appendix C of \cite{Leggett_2021}. To this sample we added the seven T dwarfs from \citet{Rothermich_2024} which were not already included in the 2021 sample. \citet{Rothermich_2024} identified these T dwarfs as wide companions to main sequence stars  and hence they are potentially important benchmark objects:
CWISE J091558.53+254713.0; CWISE J101017.43-242300.4; CWISE J104053.42-355029.7; CWISE J195956.16-644318.1; CWISE J212342.88+655615.6; CWISE J225525.22-302320.8; CWISE J233531.55+014219. We adopted the $J,H,K,$ W1,W2 (or a subset), and the parallaxes, given by \citet{Rothermich_2024} for the seven dwarfs. VIKING VISTA survey $Y$ photometry was added for CWISE J225525.22-302320.8 \citep{Edge_2013}. For  CWISE J233531.55+014219.6, where only a limit on $J$ was available, we measured a magnitude from the UKIDSS LAS survey image \citep{Lawrence_2007}, determining $J = 19.53 \pm 0.20$. AllWISE W3 was added for CWISE J104053.42-355029.7 and CWISE J233531.55+014219 (see below). Parallaxes were updated in the data compilation for all high-probability T dwarf companions identified by \citet{Rothermich_2024}, using the more precise values for the primaries. 
The updated parallaxes presented by \citet{Beiler_2024} for CWISEP J104756.81+545741.6 and CWISEP J144606.62-231717.8 were also included.
The complete dataset is given in the Appendix.

We reviewed the W3 magnitudes in our compilation  and redetermined several values. We redetermined W3 for all values with an uncertainty $\geq$0.25 magnitudes (51 sources), and we also redetermined W3 for sources  brighter than 12 with an error larger than 0.15 magnitudes (an additional 11 sources).  We redetermined W3 for the faint Y dwarf WISE J035934.06-540154.6 where the estimated uncertainty seemed too small given the faintness of the source. Finally, we examined the images for sources bright enough for an AllWISE W3 detection but with no W3 measurement reported, and determined values for three sources.

\begin{figure}
\vskip -0.2in
%\hskip -0.25in
\includegraphics[angle=-90,width = 7.3 in]{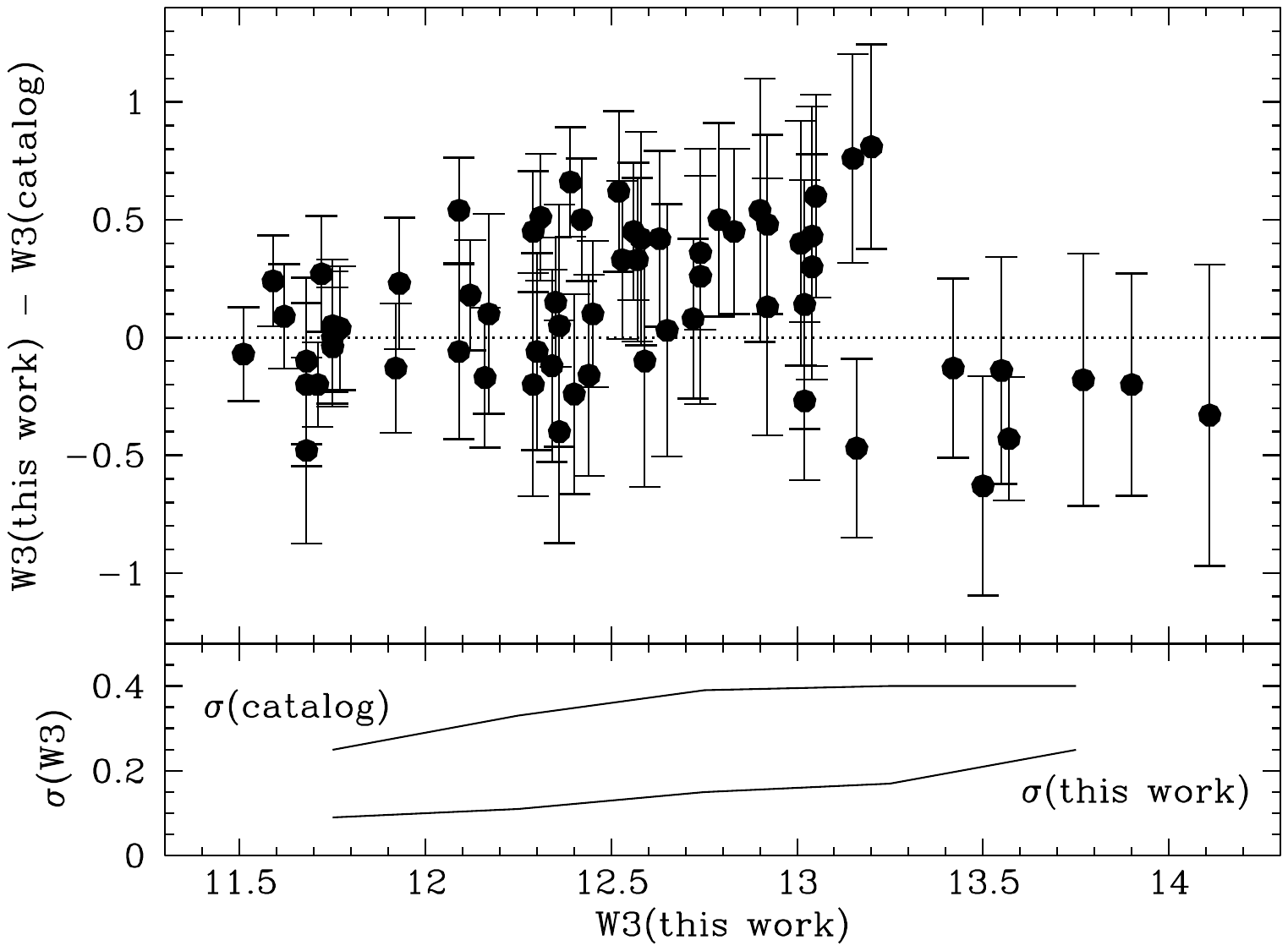}
\vskip -0.1in
\caption{The top panel shows the difference  between the previously available W3 magnitudes and those measured in this work, for 63 brown dwarfs. The lower panel shows the uncertainties in the W3 magnitudes averaged over 0.5 magnitude bins.
}
\end{figure}

W3 images were downloaded as FITS files from the IRSA {\em WISE} image service and aperture photometry was carried out; the data were calibrated using the zeropoint given in the headers. Care was taken to exclude any nearby sources, using small apertures as necessary and determining the aperture correction using bright isolated stars in the same image. Care was also taken to make sure the sky estimate was appropriate for the location of the brown dwarf. The sky estimate is the largest source of uncertainty as the {\it WISE} W3 image pixels are large, the background is high, and the background is spatially variable. The uncertainty in W3 was estimated from the scatter in the background measurements.

Table 1 gives revised W3 values for 66 brown dwarfs, and Table 2 lists 16 brown dwarfs whose previously quoted W3 magnitudes where found to be invalid on image inspection,  due to source blending or the field being too crowded for confident identification of the source.   Figure 3 shows the differences between the 63 old and new W3 magnitudes, and the estimated uncertainties as a function of W3. There is no clear trend in the differences, and the measurements presented here are significantly more precise.

%\clearpage
%\begin{longtable}
\startlongtable
\begin{deluxetable*}{lrrrr}
\tabletypesize{\normalsize}
%\tablewidth{0pt}
\tablecaption{Revised W3 Magnitudes}
\tablehead{
\colhead{AllWISE Name} & \multicolumn{2}{c}{Previous mag} & \multicolumn{2}{c}{Revised mag}  \\
\colhead{Jhhmmss.ss$\pm$ddmmss.s} & \colhead{W3} &  \colhead{err} & \colhead{W3} & \colhead{err}
}
\startdata
000517.48+373720.5 & 11.79 & 0.24 & 11.75 & 0.08 \\
001449.96+795116.1 &  13.69      &   0.40     & 13.55 & 0.27 \\
002810.59+521853.1 & 13.95 & 0.43 & 13.77 & 0.32 \\
003452.03+052306.9 &    11.78 &     0.31 &    11.68 &    0.17\\
003829.06+275852.0 & 12.38 & 0.33 & 12.83 & 0.12 \\
004143.14-401924.3 & 12.36 &     0.39 &    12.30 &    0.15\\
004945.61+215120.0 & 11.58 & 0.19 & 11.51 & 0.06 \\
005021.05-332228.8 & 11.84 & 0.24 & 12.29 & 0.09 \\
010650.61+225159.1 & 12.49 & 0.46 & 12.29 & 0.11 \\
013217.78-581825.9&    14.10 &     0.40 &    13.90 &    0.25\\
014603.23-261908.7	& 13.63 & 0.34 & 13.16 & 0.17 \\
014837.52-104803.9 &     12.69 &    0.52 & 12.59 &    0.12\\
030919.70-501614.2AB&    12.76 &     0.46 &    12.36 &    0.11\\
031326.00+780744.3 & 12.05 & 0.26 & 11.92 & 0.09 \\
032547.73+083118.2 & 11.88 &0.34 & 11.68 & 0.07 \\
035000.31-565830.5 & 12.33 & 0.28 & 12.16 & 0.10 \\
035934.06-540154.6 & 14.00   &     0.20 &  13.57  &    0.17 \\
041022.75+150247.9 & 12.31 &    0.50 &    12.36&    0.12\\
041102.17+471423.6 & 12.36 & 0.50& 12.90 & 0.25 \\
062720.07-111428.0 & 11.53 & 0.21 & 11.62 & 0.07 \\
075108.79-763449.6 &    11.91 &     0.16 &  11.71  &    0.08 \\
081117.95-805141.4 & 12.64 & 0.32 & 12.72 & 0.11 \\
094306.00+360723.3 & 12.29 & 0.39 & 12.79 & 0.13 \\
095047.31+011733.1 &  &     &   12.50 &    0.12 \\
104053.42-355029.7&    12.62 &     0.52 &    12.65 &    0.13\\
105349.41-460241.2& 14.13 & 0.40 & 13.50 & 0.24 \\
105512.93+544329.7 &   11.55 &     0.20 &  12.09  &    0.10\\
111448.74-261827.9 & 11.35 & 0.18 & 11.59 & 0.07 \\
111838.69+312537.7 & 12.24  &     0.33 &   12.57 &    0.11\\
114156.67-332635.5& 11.73 & 0.21 & 12.39 & 0.10\\
115013.85+630241.3 & 12.38 & 0.30 & 12.74 & 0.13 \\
115229.63+035926.8	& 12.48 & 0.50 & 12.74 & 0.21 \\
121710.27-031112.1 & 11.70 & 0.26 & 11.93 & 0.10\\
123738.95+652620.3 & 11.94 & 0.22 & 12.12 & 0.08 \\
125721.01+715349.3&  13.55  &  0.33   &    13.42 &    0.19\\
125804.91-441232.6 & 12.11 & 0.27 & 12.56 &0.11 \\
130041.63+122114.5 & 11.70 & 0.27 & 11.75 & 0.08\\
130217.08+130851.0&    12.16 &     0.42 &    12.58 &    0.17 \\
133553.41+113004.7 & 12.15 & 0.36 &12.09  & 0.09\\
134646.07-003151.4 & 11.92 & 0.24 & 12.42 & 0.10 \\
140518.32+553421.3 & 12.20 & 0.26 & 12.35 & 0.09\\
144901.84+114710.7 & 12.74 & 0.45 & 13.04 & 0.16 \\
145731.67+472420.1&    12.64 &     0.41 &    12.40 &    0.11\\
150411.80+102715.6 &    12.60 &     0.41 &    12.44 &    0.12\\
150457.56+053759.8 &    &     &  12.65  &    0.12 \\
154151.70-225019.1 & 12.20 & 0.30 & 12.53 & 0.15 \\
162414.07+002915.6& 12.16 & 0.39 & 11.68 & 0.06 \\
162838.09+230822.7 & 11.80 & 0.25 & 12.31 & 0.10 \\
170745.84-174452.4 & 12.07 &     0.41 &    12.17 &    0.11 \\
173835.52+273258.8 & 12.45 & 0.40 & 13.05 & 0.16 \\
175805.45+463316.9 & 12.35 & 0.29 & 12.45 & 0.11 \\
181006.18-101000. 5&   11.45 &    0.22 &    11.72 &    0.11\\
182831.08+265037.6 & 12.44 & 0.34 & 12.92 & 0.17 \\
183207.94-540943.3 &  12.61  &     0.52 &    13.04 &    0.18 \\
193054.55-205949.4 & 14.44 & 0.58 & 14.11 & 0.27 \\
200520.35+542433.6 &   12.39 &     0.40 &   13.20 &    0.17 \\
205628.91+145953.2 & 11.73 & 0.25 & 11.77 & 0.08 \\
213456.79-713744.7 & 12.39 & 0.40 & 13.15 & 0.19 \\
220905.75+271143.6 & 12.46 & 0.39 & 12.34 & 0.12\\
222829.01-431029.8 & 11.75 & 0.27 & 11.75 & 0.08\\
223204.53-573010.4 &    11.90 &    0.32 &  12.52  &    0.12 \\
225404.16-265257.5 & 13.29 & 0.29 & 13.02 & 0.17 \\
230158.29-645857.5 & 12.61 & 0.48 & 13.01 & 0.20 \\
233531.55+014219.6 &    &     &    12.81 &    0.20\\
234351.20-741846.9 &    12.79 &     0.50 &    12.92 &    0.22\\
235120.61-700025.7 & 12.88 & 0.51 & 13.02 & 0.14 \\
\enddata
\end{deluxetable*}
%\end{longtable}

%\clearpage
\begin{deluxetable*}{lll}[!hb]
\tabletypesize{\normalsize}
%\tablewidth{0pt}
\tablecaption{Excluded W3 magnitudes Due to Source Confusion or Blending}
\tablehead{
\colhead{AllWISE Name} & 
\colhead{AllWISE Name} & 
\colhead{AllWISE Name} 
}
\startdata
003231.09-494651.4 & 080622.22-082046.4 &  145838.12+173447.7  \\
003507.79-153230.7 &   083019.97-632305.4 & 181849.69-470146.4\\
004158.29+381811.9 &  094005.50+523359.2 &  214025.23-332707.4\\
032517.68-385453.8 &  132233.63-234017.0 & 235716.49+122741.5  \\
071322.55-291752.0 &  141127.88-481151.5 & \\
072719.15+170951.3 &   141623.96+134836.0 &  \\
% 001449.96+795116.1 %allwise no w3?
% 002810.59+521853.1	%allwise no w3?
% 193054.55-205949.4 %allwise no w3
% 105349.41-460241.2 %allwise no w3?
\enddata
\end{deluxetable*}

\bigskip
\section{Analysis}

\subsection{Models}

Over the last 20 years significant improvements have been made in modelling the atmospheres of cool stars and brown dwarfs. These include more complete molecular line lists \citep[e.g.][]{Saumon_2012, Yurchenko_2014, Polyansky_2018}, a better understanding of vertical mixing and the resulting disequilibrium chemistry \citep[e.g.][]{Saumon_2006, Zahnle_2014}, and the incorporation of grain formation and sedimentation \citep[e.g.][]{Ackerman_2001,  Morley_2012, Morley_2014, Lacy_2023}.  However, while synthetic spectra and photometry reproduced the observations of cool stars and brown dwarfs with $T_{\rm eff} > 600$~K, systematic discrepancies of $\approx 50\%$ existed at wavelengths around $1.2~\mu$m and at 2 to 4$~\mu$m, for cooler brown dwarfs \citep[e.g.][]{Leggett_2017, Karalidi_2021, Lacy_2023}; the modelled near-infrared flux was typically too bright and the 2 to 4$~\mu$m flux was typically too faint.

For cold brown dwarfs, the near-infrared flux emerges from layers deep in the atmosphere, and the 2 to 4$~\mu$m flux emerges from higher layers \citep[e.g.][their Figure 17]{Karalidi_2021}. Knowing this, \citet{Leggett_2021} explored modifications to the standard radiative/convective, adiabatic, pressure-temperature relationship, in order to make the lower atmosphere colder and the upper atmosphere warmer. The atmospheric structure was parameterized by a 
pressure $P_{(\gamma, max)}$,  defining the atmospheric depth  above which the adiabatic parameter $\gamma$ is less than the standard value ($\approx 1.4$ for hydrogen). Tuning the models to reproduce the $1 \lesssim \lambda~\mu$m $\lesssim 20$ energy distributions of one T9 dwarf and six Y dwarfs, values of $\gamma$ between 1.20 and 1.33 were derived, with $7 \lesssim P_{(\gamma, max)}$~bar $\lesssim 50$. Using the modified adiabat improved agreement with spectroscopic and photometric observations of cold brown dwarfs by factors of two to five at 
$1 \lesssim \lambda~\mu$m $\lesssim 5$ \citep[][Figures 12 and 13]{Leggett_2021}.

The $P_{(\gamma, max)}$ layer for the \citet{Leggett_2021} sample corresponded to temperatures around 800~K, a region where the gaseous nitrogen chemistry is changing and chlorides and sulfides are condensing \citep{Lodders_1999, Morley_2012, Leggett_2021}, possibly disrupting convection and leading to a diabatic profile. The rapid rotation of the cool brown dwarfs \citep{Zapatero_2006, Tannock_2021} is also expected to alter the pressure-temperature profile of the atmosphere \citep[e.g.][their Figure 4]{Tan_2021}. Retrieval analyses have determined
similar changes to the atmospheric profile, see \citet{Kothari_2024, Lew_2024} for recent examples using {\it JWST} data.

\citet{Leggett_2021} generate a grid of models which include disequilibrium chemistry and adopt a diabatic profile with an effective adiabatic parameter of $\gamma = 1.25$ and $P_{(\gamma, max)} = 15$~bar; these are known as 
the ATMO 2020++ model atmospheres.  Note that the adopted values for $\gamma$ and $P_{(\gamma, max)}$  are empirical and based on the fit to the energy distributions, they have not been quantitatively connected to a diabatic or kinetic process.

The ATMO 2020++ grid was expanded to include a wider range in metallicity for \citet{Meisner_2023}. Later, a grid of phosphine-free ATMO 2020++ models was generated for  \citet{Leggett_2023, Leggett_2024}  because {\em JWST} data show it to be absent \citep{Beiler_2023, Leggett_2023, Lew_2024, Luhman_2024}. Phosphine is a chemical disequilibrium species which is expected to be the stable form of phosphorus in cool atmospheres
\citep{Visscher_2006}; although the feature is seen in solar system giant planets, it may be that the different composition or gravity of the brown dwarf atmosphere results in phosphorus taking a different form.
ATMO 2020++ synthetic spectra and colors are available on the ERC ATMO OpenData web site\footnote{\url{https://opendata.erc-atmo.eu}}.

We use the phosphine-free ATMO 2020++ models in this work. The adiabat is modified at atmospheric pressures between 0.15 and 15 bars at log ~$g = 4.5$, which are scaled by $\times 10^{({\rm log~} g - 4.5)}$ at other surface gravities. Out-of-equilibrium chemistry is used with $K_{zz} = 10^5$ cm$^2$ s$^{-1}$ at log ~$g = 5.0$, which is scaled by
$\times 10^{2(5 - {\rm log} ~g)}$ at other surface gravities. 
The model grid covers $250 \leq T_{\rm eff}$~K $\leq 1200$ (250~K, 275~K, 300~K, 350~K, 400~K, 450~K, 500K, then every 100~K to 1200~K), 
for three metallicities:  [m/H] $= -0.5, 0, +0.3$.

Our brown dwarf sample is local -- 90\% of the brown dwarfs with measured distances lie within 25~pc -- and 
a range in metallicity between $-0.5$ and $+0.3$ covers the likely range of values for the population \citep[e.g.][their Figure 4]{Hinkel_2014}. 
Evolutionary models calculate that brown dwarfs with $T_{\rm eff}$ values between 300~K and 800~K, and a likely age range of 0.5~Gyr to 6~Gyr \citep{Dupuy_2017,
Kirkpatrick_2021a, Best_2024}, have a range in log $g$ of 3.5 to 5.0 \citep{Marley_2021}.
The metal-poor grid includes log $g$ values of 4.0, 4.5 and 5.0; the solar metallicity grid includes log $g$ values of 3.5, 4.0, 4.5 and 5.0; and the metal-rich grid includes log $g$ values of 2.5 to 5.5 in steps of 0.5~dex.   

\subsection{Sensitivity to Metallicity and Gravity}

Figure 2 shows a synthetic 450~K spectrum for a brown dwarf at 10~pc, with radii determined from evolutionary models for each $T_{\rm eff}$ and gravity \citep{Marley_2021}. {\it JWST} and {\it WISE} filter profiles are overlaid. The spectrum is also shown divided by spectra with the same $T_{\rm eff}$ but with lower gravity (middle panel) and lower metallicity (bottom panel).  The radius of the log $g = 4.0$ brown dwarf is $1.35 \times$ larger than the log $g = 5.0$ brown dwarf, meaning an increase in brightness of 80\%. Without the radius adjustment the red comparison line in the middle panel of Figure 2 would lie at $y \approx 0.6$.

Figure 2 shows that the near-infrared region is very sensitive to both metallicity and gravity, as previously noted for T dwarfs \citep[e.g.][]{Burgasser_2006b,Liu_2007} and Y dwarfs \citep[e.g.][]{Leggett_2017}.
The $10 \lesssim \lambda~\mu$m $\lesssim 15$ region shows some sensitivity to metallicity and gravity but not to the same degree.  Interestingly, the average flux over the W2 (or F444W) bandpass is approximately equal in all three cases. The absorption features at these wavelengths (primarily due to CO and CO$_2$) increase with decreasing gravity, however the increase in radius compensates for the additional absorption, at this temperature. 

Figure 4 illustrates the gravity and metallicity sensitivity for cold brown dwarfs using the W2 $-$ W3 color, with $M_{W2}$ or $J -$ W2 as a proxy for $T_{\rm eff}$. As $T_{\rm eff}$ decreases below 600~K the intrinsic spread of the W2 $-$ W3 color increases significantly, primarily caused by an increase in sensitivity to gravity. For Y dwarfs, with 
$T_{\rm eff} \lesssim 500$~K, the W2 $-$ W3 color increases by 0.3 to 0.6 magnitudes for a 0.5 dex decrease in log $g$. A 0.5 dex decrease in [m/H] results in smaller changes of 0.1 to 0.3 magnitudes. Brown dwarfs (or exoplanets) with very blue or very red W2 $-$ W3 colors are identified in the figure and discussed in detail below.

Note also that for Y dwarfs warmer than $\sim 400$~K, $M_{W2}$ changes very little ($\sim 20\%$) as gravity and metallicity are varied and $T_{\rm eff}$ kept constant. Hence $M_{W2}$ (or a similar bandpass) can be a useful luminosity indicator.

%\clearpage
\begin{figure}
\vskip -0.4in
\hskip 0.5in
\includegraphics[width = 5.0 in]
{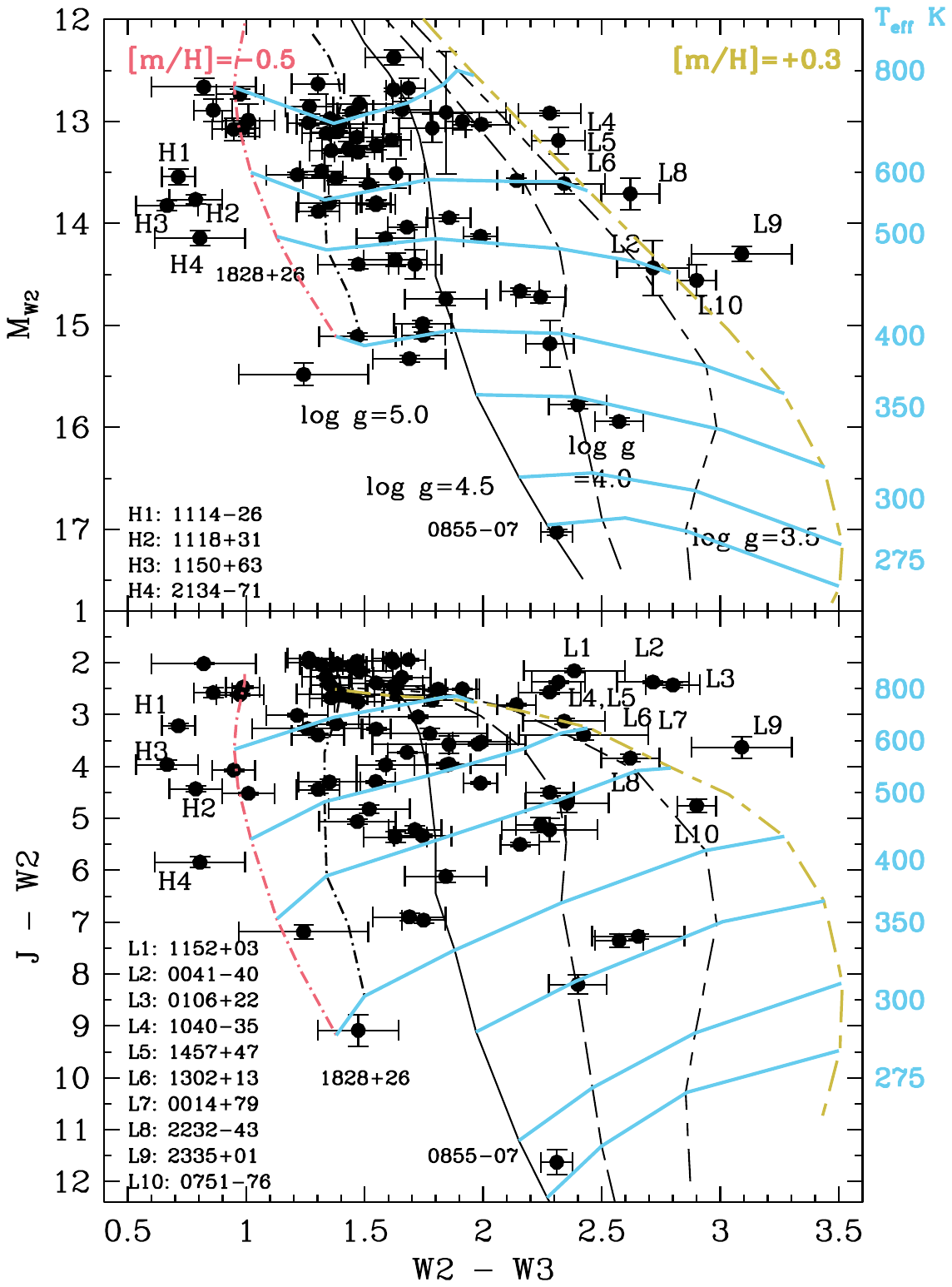}
\vskip -0.3in
%\hskip 0.5in
\caption{Color magnitude diagrams for late-T and Y dwarfs. Black circles are data points from the compilation presented here. Sources with extreme W2 $-$ W3 colors are identified by an abbreviated {\it WISE} RA-Decl.. The locations of the unusual Y dwarf WISEPA J182831.08+265037.8, and the very cold WISE J085510.83-071442.5, are also indicated. 
Black almost vertical lines are solar metallicity ATMO 2020++ PH$_3$-free sequences for log $g =$ 3.5, 4.0, 4.5, and 5.0 from right to left. The red line is a log $g =$ 5.0 sequence with [m/H] $= -0.5$ and the yellow line is 
a log $g =$ 3.5 sequence with [m/H] $= +0.3$. 
The cyan lines are isotherms, with $T_{\rm eff}$ indicated along the right axes. At these temperatures, log $g = 3.5$ corresponds to a mass of $\approx 3 M_{Jup}$, 4.0 to $\approx 6 M_{Jup}$, 4.5 to $\approx 14 M_{Jup}$, and 5.0 to $\approx 35 M_{Jup}$ \citep{Marley_2021}.  Temperature decreases with age for a given mass; an age of 15~Gyr corresponds to $T_{\rm eff}$ values of 400~K and 260~K for log $g =$ 5.0 and 4.5 respectively, and colder values for lower gravities \citep{Marley_2021}.
}
\end{figure}

\subsection{Candidate Young, Very Low Mass Brown Dwarfs or Exoplanets}

Ten sources are identified in Figure 4 which lie along or redwards of the log $g = 3.5$ sequences. Their $M_{W2}$ and $J -$ W2 colors indicate $500 \lesssim T_{\rm eff}$~K $\lesssim 800$.  Evolutionary models then imply young ages of 10 to 80~Myr and a low mass of around 3~$M_{\rm Jup}$ \citep[see e.g. Figure 10 of][]{Marley_2021}. These young objects may also be metal-rich, however a metallicity $>>+0.3$~dex is unlikely for
this local sample \citep[e.g.][]{Nordstrom_2004,Rojas_2012, Hinkel_2014}.   Each source is discussed below, and a summary of their properties is given in Table 3.

\begin{deluxetable*}{llcc}[!hb]
\tabletypesize{\normalsize}
%\tablewidth{0pt}
\tablecaption{Estimated Properties of Candidate Young $3~M_{\rm Jup}$ Objects}
\tablehead{
\colhead{AllWISE Name} & \colhead{Other Name(s)} & \colhead{$T_{\rm eff}$~K}  &
\colhead{Age~Myr}
}
\startdata
001449.85+795115.9$^a$ &  & 600 & 10 -- 100  \\
004143.77-401929.9 & & 500 -- 800 &  10 -- 40 \\
010650.61+225159.1 &  & 800 &  4 -- 10 \\
075108.79-763449.6$^b$  & COCONUTS-2b, L 34-26b & 450 & 30 -- 80  \\ 
104053.42-355029.7 & UPM J1040-3551b & 700 &  10 -- 30  \\
115229.64+035926.8   &  ULAS J115229.67+035927.2 & 800 &  5 -- 15  \\
130217.21+130851.2$^a$  &  & 600 &  20 -- 40 \\
145731.67+472420.1$^c$ & PSO J224.3820+47.40 & 700 &  10 -- 30 \\
223204.50-573010.5 &   & 550  & 15 -- 40 \\
233531.55+014219.6$^{a,d}$  &  GJ 900b & 500 & 15 -- 40  \\
\enddata
\tablecomments{\\
$T_{\rm eff}$ is estimated from the isotherms shown in Figure 4; age is estimated from evolutionary models \citep{Marley_2021}, adopting this  $T_{\rm eff}$ and log $g \approx 3.5$, and taking into account the error bars shown in Figure 4. Estimated uncertainties are $\sim 50$~K and $\sim 50\%$ in age (however see other age estimates in the following notes).\\
$^a$ Kinematics suggest membership of the Carina-Near moving
group with an age of $\sim 200$ Myr \citep{Gagne_2018}.\\
$^b$ \citet{Zhang_2021a} estimate an age of 150 –- 800~Myr using the M dwarf primary; for the brown dwarf they use a luminosity estimate to determine: $T_{\rm eff} = 434$K, log $g=4.11$, mass $= 6.3~M_{\rm Jup}$.\\
$^c$ Near-infrared spectral analyses by \citet{Zhang_2021b} and \citet{Zalesky_2022} find $T_{\rm eff} \approx 900$~K, log $g \approx 3.7$, implying a mass and age of $\sim 4~M_{\rm Jup}$ and $\sim 15$~Myr from evolutionary models \citep{Marley_2021}.\\
$^d$ A companion to the triple  system GJ 900, with an age based on rotational measurements of 100 -- 300~Myr
\citep{Rothermich_2024}.}
\end{deluxetable*}

Three of the objects have kinematics consistent with membership of the Carina-Near moving group with an age of $\sim 200$~Myr \citep{Gagne_2018}: WISE J001449.85+795115.9, T8; ULAS J130217.21+130851.2, T8.5; CWISE J233531.55+014219.6, T9. A parallax measurement would strengthen the case for group membership for WISE J001449.85+795115.9, where the 77\% likelihood is based on location and proper motion only. \citet{Zhang_2021b} report that ULAS J130217.21+130851.2 
has much redder $J - K$ and $H - K$ colors than other T8 -- T9 field dwarfs, and has slightly enhanced fluxes near the $Y$-band peak, suggesting a lower surface gravity. \citet{Rothermich_2024} report that CWISE J233531.55+014219.6 is a companion to the triple K and M dwarf system GJ 900, and that the age of the system (based on rotational measurements) is 100 -- 300~Myr. 

The T9 WISEP J075108.79-763449.6, also known as COCONUTS-2b or L 34-26b \citep{Zhang_2021a}, is the coolest source. Although its kinematics do not associate it with any young moving group, the tangential velocity is low at $11$ km s$^{-1}$, supporting classification as a young field dwarf  \citep{Zhang_2021a}. \citet{Zhang_2021a}  estimate an age for the solar-metallicity M dwarf primary using its spectroscopic, activity, rotation, kinematic, and photometric properties; they determine an age of 150 –- 800~Myr for the system. Those authors estimate the luminosity of the secondary from its near-infrared magnitudes, and combine this with the age estimate to constrain the temperature, gravity and mass of the secondary to $T_{\rm eff} = 434 \pm 9 $K, log $g=4.11^{+0.11}_{-0.18}$, mass $= 6.3^{+1.5}_{-1.9}~M_{\rm Jup}$, from both hot- and cold-start evolutionary models. These properties are consistent with the location of the source in Figure 4.  WISEP J075108.79-763449.6 has been observed by {\it JWST} as part of program 3514, PI Mickael Bonnefoy, and therefore the properties of this interesting source should soon be better known.

WISE J145731.67+472420.1 (also PSO J224.3820+47.4057) is classified as a T7 dwarf by \cite{Best_2015}. Near-infrared spectral analyses appear to confirm the low gravity indicated in Figure 4.  Forward modelling by \citet{Zhang_2021b} assigns $T_{\rm eff} = 921 \pm 4 $K and log $g = 3.67^{+0.32}_{-0.01}$, while a retrieval analysis by \citet{Zalesky_2022} finds $T_{\rm eff} = 893^{+9}_{-26}$K, log $g = 3.65^{+0.25}_{-0.14}$. Evolutionary models \citep{Marley_2021} then provide mass and age estimates of $\sim 4~M_{\rm Jup}$ and $\sim 15$~Myr.
The parallax and proper motion \citep{Best_2020} give a low tangential velocity of 16~km s$^{-1}$, supporting a young age \citep[e.g.][their Figure 31]{Dupuy_2012}, although the BANYAN tool \citep{Gagne_2018} assigns field membership.  

The remaining five young planetary-mass candidates require additional observations to confirm their nature:
\begin{itemize}
    \item CWISE J004143.77-401929.9 is classified as a T8-peculiar
    by \citet{Kirkpatrick_2021a}, and it has an enhanced $K$-band flux, suggestive either of low gravity or high metallicity, or both.  The BANYAN tool \citep{Gagne_2018} estimates field membership for this source, that is, not young, and the tangential velocity is high at $95$ km s$^{-1}$, which would also not support a young age \citep[e.g.][their Figure 31]{Dupuy_2012}. However the location of the source in the two panels of Figure 4 is inconsistent. Agreement would be improved if the trigonometric parallax is underestimated and the target is closer than inferred from the parallax; this would also reduce the tangential velocity. An improved parallax measurement and {\it JWST} observations would help constrain the nature of this source.
\item CWISEP J010650.61+225159.1 is classified as a T6.5 dwarf by \citet{Meisner_2020a}. There is no published trigonometric parallax or near-infrared spectrum. Both of these would be useful. The BANYAN tool \citep{Gagne_2018} estimates field membership for this source based on location and proper motion alone.
\item 
CWISE J104053.42-355029.7 is classified as a T7 dwarf by \citet{Rothermich_2024}, who also determine that it is a wide companion to the M dwarf UPM J1040-3551 (or 2MASS J10405549-3551311), which has a metallicity close to solar ([Fe/H] $\approx-0.06$). The tangential velocity is low at 15~km s$^{-1}$, possibly supporting a young age \citep[e.g.][their Figure 31]{Dupuy_2012}, although the BANYAN tool \citep{Gagne_2018} assigns a field membership. Further studies of the M dwarf primary, to constrain the age of the system, would be valuable.
\item ULAS J115229.67+035927.2 is classified as a T6 by \citet{Scholz_2012}; the near-infrared spectrum appears typical for its type. The proper motion of the source is very uncertain, and there is no published trigonometric parallax  measurement; the BANYAN tool \citep{Gagne_2018} is unable to constrain moving group membership.  Additional astrometry and a parallax measurement would be helpful for this source.
\item WISE J223204.50-573010.5 is classified as a T9 dwarf by \citet{Tinney_2018}, and the tangential velocity \citep{Kirkpatrick_2019} of 39~km s$^{-1}$, suggests thin disk membership \citep[e.g.][their Figure 31]{Dupuy_2012}; the BANYAN tool \citep{Gagne_2018} assigns field membership. A near-infrared spectrum covering the $K$-band would be helpful, as would {\it JWST} data. 
\end{itemize}

\subsection{Candidate Old and  Metal-Poor Brown Dwarfs}

Four sources are identified in Figure 4 with W2 $-$ W3 $\approx 0.7$ and $14.2 \geq M_{\rm W2} \geq 13.5$, implying $500 \leq T_{\rm eff}$~K $\leq 600$. The models indicate that they have log $g > 5$ and are significantly metal-poor. Evolutionary models then imply an age $\geq 8$~Gyr and a mass $\geq 35~M_{\rm Jup}$ .

One of these sources, WISE J111838.70+312537.9, is a T8.5
member of the $\xi$ Ursae Majoris system \citep{Wright_2013}. \citet{Wright_2013}, and references therein, report [Fe/H] $= -0.32$ for the system and an age between 2 and 8~Gyr. These are consistent with the model inferences above, if the age is towards the higher limit.

The other three sources are known metal-poor brown dwarfs, from studies by \citet{Zhang_2019, Zhang_2021b} using near-infrared spectroscopy and model analyses: 
WISEA J111448.74-261827.9, T7.5; WISE J115013.85+630241.5, T8; and WISEP J213456.73-713743.6, T9pec. \citet{Zhang_2021b} estimate [m/H] $\approx -0.4$ for WISEA J111448.74-261827.9, a value which appears typical for the T dwarfs in their samples. These sources all have a tangential velocity of $\lesssim 60$~km s$^{-1}$ suggesting thin disk membership \citep[e.g.][their Figure 31]{Dupuy_2012} and an age not significantly higher than 8~Gyr \citep[e.g.][]{Haywood_2013}.

\subsection{The Unusual Y Dwarf WISEPA J182831.08+265037.8}

The Y dwarf WISEPA J182831.08+265037.8 (hereafter WISE 1828) has been recognized to be unusual since its discovery by \citet{Cushing_2011}. Historically, it was a challenge to reproduce the observations using the then available models \citep[e.g.][]{Beichman_2013}. Also, although its super-luminosity suggests multiplicity, {\it JWST} images exclude a companion at distances $> 0.5$~au \citep{Furio_2023}. In our previous analysis using ATMO 2020++ models, we estimated that the system is likely to be an identical pair of brown dwarfs with $T_{\rm eff} = 375$~K, log $g = 4.0$, and [m/H] $= -0.5$ \citep{Leggett_2021}. That fit was based on a near-infrared spectrum and mid-infrared photometry.

In this work we revise the W3 photometry for WISE 1828 (Section 2, Table 1). The bluer W2 $-$ W3 color places this dwarf on the log $g = 5$ model sequence in Figure 4. While the $M_{\rm W2}$:W2$-$W3 plot (upper panel of Figure 4) suggests it could be a single 500~K brown dwarf, the $J -$ W2:W2$-$W3 plot (lower panel) implies $T_{\rm eff} \approx 400$~K; the colors are consistent if the source is a 400~K binary composed of near-identical dwarfs. 
Moderate resolution mid-infrared spectra are now available for WISE 1828, and we can use these data to further explore the use of the W2 $-$ W3 color as a gravity indicator.  \citet{Lew_2024} present 2.9 to 5.1~$\mu$m NIRSpec spectra, and \citet{Barrado_2023} present 4.9 to 18~$\mu$m MIRI spectra. 
Figure 5 compares the observed NIRSpec and MIRI spectra to various ATMO 2020++  PH$_3$-free synthetic spectra. 
We use the measured distance to the dwarf, and constrain the radius to be equal to the evolutionary model values \citep{Marley_2021}, to scale the synthetic spectra to the observations.

\begin{figure}
\vskip -0.1in
%\hskip 0.4in
\includegraphics[angle=-90,width = 7.3 in]
{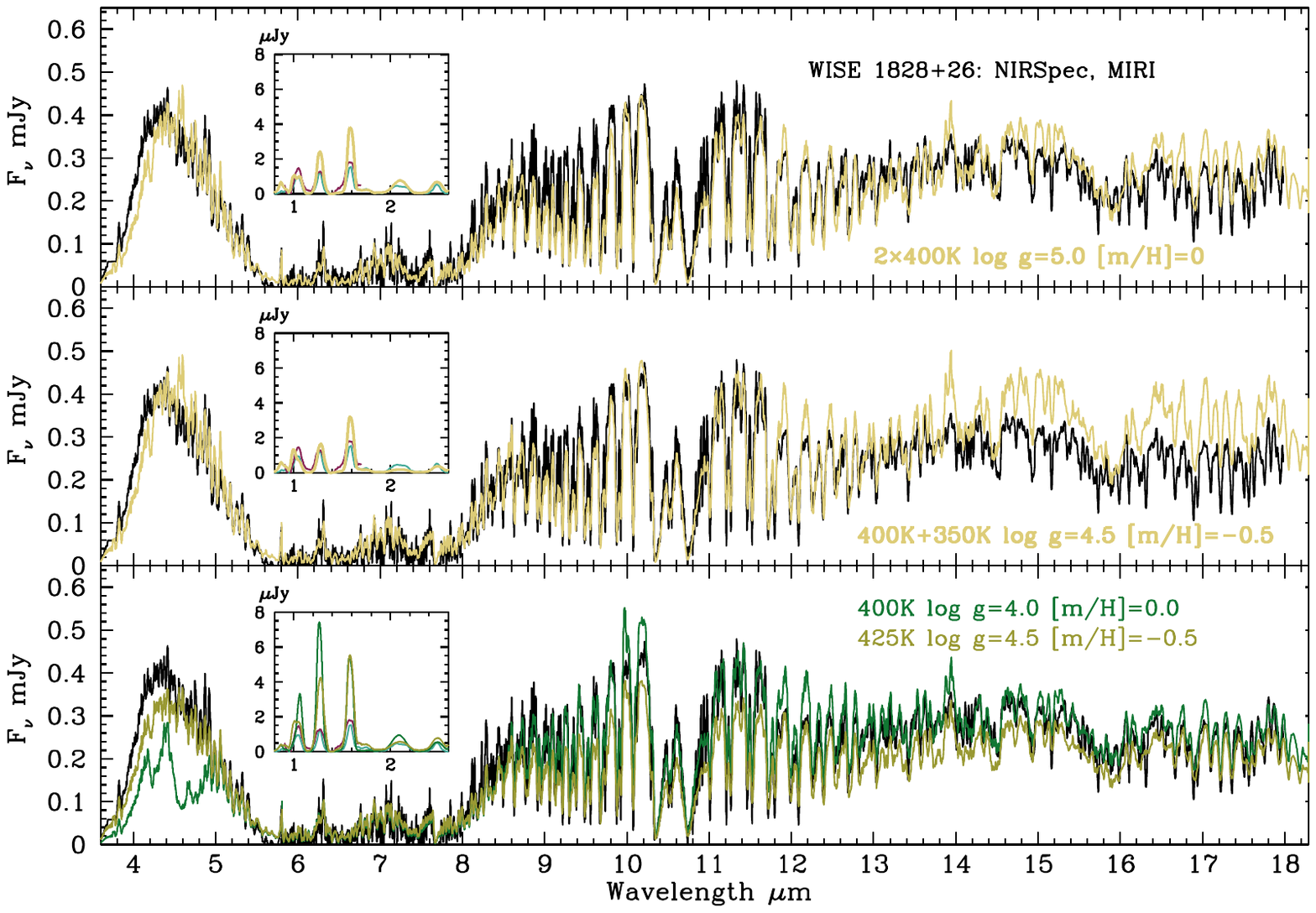}
%\vskip -0.4in
\caption{The black lines are {\it JWST} NIRSpec and MIRI data for WISEPA J182831.08+265037.8, from \citet{Barrado_2023} and \citet{Lew_2024}. The insets show the near-infrared region, where cyan lines are the low-resolution NIRSpec data ({\it JWST} Cycle 1 GTO program 1189, PI Thomas Roellig) and red lines are {\it Hubble Space Telescope} data from \citet{Cushing_2021}. 
Yellow, olive and green lines are ATMO 2020++ PH$_3$-free synthetic spectra, with parameters given in the legends; yellow spectra adopt a binary solution, the olive and green spectra are for a single dwarf.  Scaling the synthetic spectra is done by adopting the brown dwarf radius as calculated by evolutionary models \citep{Marley_2021} for each $T_{\rm eff}$ and log $g$, and using the measured distance to the Y dwarf. The upper two panels reproduce the observations quite well, with the top spectrum being our preferred solution. In the bottom panel, the olive line is an ATMO 2020++ synthetic spectrum with parameters similar to the \citet{Lew_2024} Elf-Owl fit to the $2.9 \lesssim \lambda~\mu$m $\lesssim 5.1$ NIRSpec spectrum.  The green line is the 
 ATMO 2020++ synthetic spectrum adopted by
\citet{Barrado_2023} as the best-fit ATMO model to their $5.0 \lesssim \lambda~\mu$m $\lesssim 18$ MIRI spectrum. See text for further discussion.
}
\end{figure}

We explored binary solutions to the observed NIRSpec and MIRI spectra, given that the $J$, W2, W3 color-set can only be reproduced if the system is a pair of near-identical 400~K dwarfs (Figure 4). The colors are matched by ATMO 2020++ model atmospheres with log $g =$ 4.5 or 5.0, solar or sub-solar metallicity, and $350 < T_{\rm eff}$~K $< 450$ (Figure 4);  we restricted comparison spectra to those parameter ranges, as available in our grid (Section 3.1).

We interpolated the observations and the synthetic spectra to a sampling frequency of $\delta \lambda = 0.0025~\mu$m and calculated Goodness-of-Fit (GF) statistics for each solution by summing the squares of the difference between the observed and modelled flux, divided by the uncertainty in the observed flux.  We explored using different wavelength regions for the GF calculation, avoiding wavelengths where there is very little signal. The best fits were found for the two binary solutions shown in the top panels of Figure 5:
(1) a pair of 400~K, log $g = 5.0$, [m/H] $= 0$ brown dwarfs, and (2) a system composed of a 400~K and a 350~K brown dwarf, both with log $g = 4.5$ and  [m/H] $= -0.5$. 
The typical average deviation is $5~\sigma$, although both fits underestimate the flux at $3.8 \lesssim \lambda~\mu$m $\lesssim 4.2$ (Figure 5). Most likely this is due to the known remaining deficiencies in the models, which calculate too low a flux around the $3.3~\mu$m CH$_4$ absorption band \citep[see discussion in][]{Leggett_2021}. The two solutions have similar GF values, with a slight preference for the solar metallicity, higher gravity, solution (for example, for $4.3 \leq \lambda~\mu$m $\leq 17.9$ the average deviation is $5.1\sigma$ for solution (1) and $5.7\sigma$ for solution (2)). Figure 6 shows a more detailed comparison using solution (1). The 
near-infrared fit (seen in the inset panel of  Figure 5), could be improved by 
tuning $\gamma$ and $P_{(\gamma, max)}$ to adjust the pressure-temperature profile in the interior of the atmosphere, where these fluxes originate \citep[Figure 6 of][]{Leggett_2021}.

\begin{figure}
\vskip -0.2in
%\hskip 0.4in
\includegraphics[angle=-90,width = 7.3 in]
{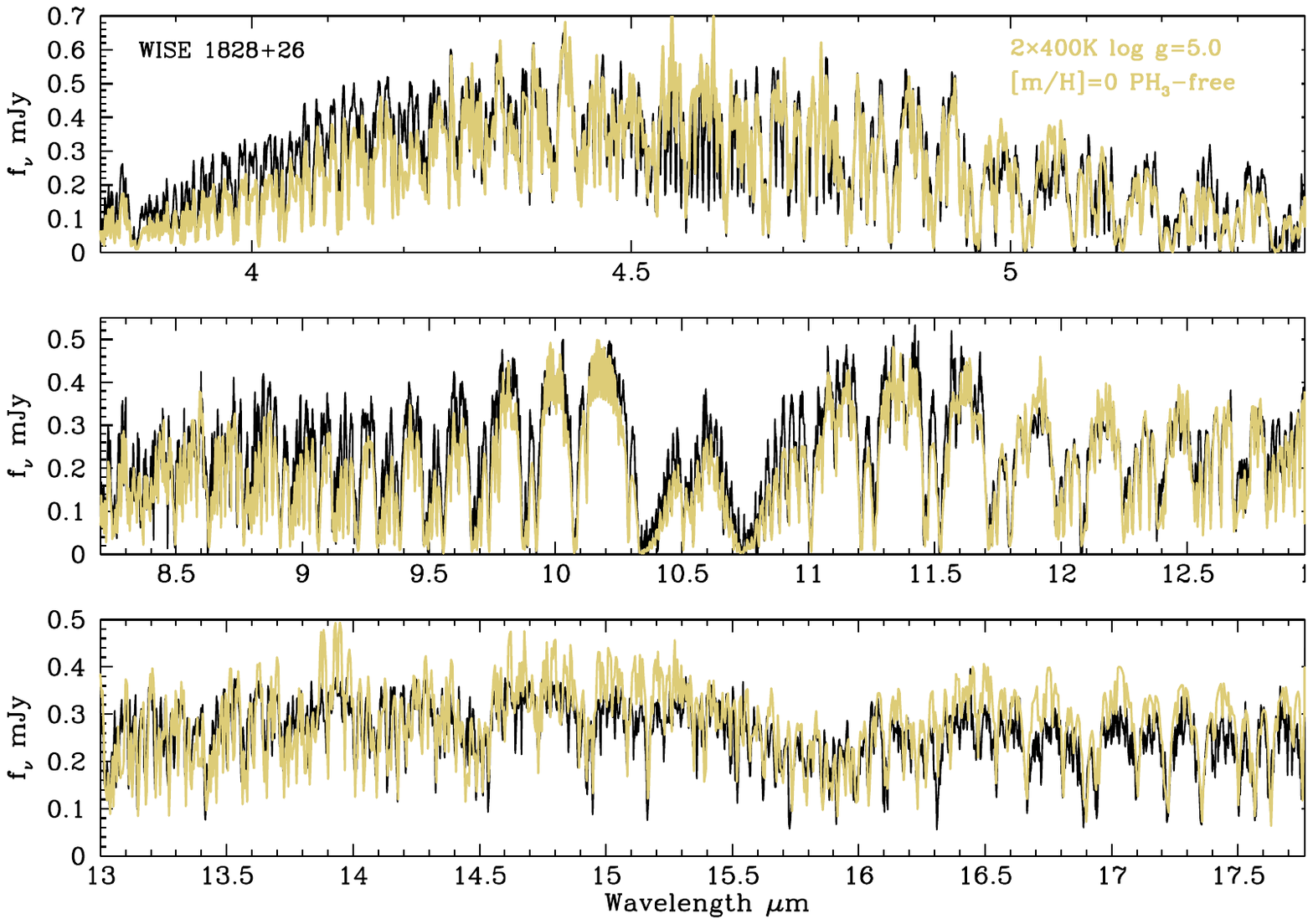}
%\vskip -0.4in
\caption{A more detailed comparison of the {\it JWST} NIRSpec and MIRI data for WISEPA J182831.08+265037.8 \citep{Barrado_2023, Lew_2024} and the synthetic ATMO 2020++ spectrum for the adopted solution for the system, an identical pair of 400~K log $g = 5.0$ solar metallicity dwarfs. Absorption features due to (primarily) CO, CO$_2$, CH$_4$, H$_2$O, and NH$_3$ are identified in Figure 5 of \citet{Lew_2024} and Figure 1 of \citet{Barrado_2023}.
}
\end{figure}

Figure 5 also shows comparisons of the data to synthetic ATMO 2020++ spectra with atmospheric parameters similar to those found in model analyses of the NIRSpec and MIRI spectra separately, by \citet{Lew_2024} and \citet{Barrado_2023}. 
\begin{itemize}
    \item 
\citet{Lew_2024} used Elf-Owl models \citep{Mukherjee_2022} with the $2.9 \lesssim \lambda~\mu$m $\lesssim 5.1$ NIRSpec data to determine 
$T_{\rm eff} = 425~K$, log $g = 4.4$, and [m/H] $= -0.6$. The bottom panel of Figure 5 shows the ATMO 2020++ spectrum for similar parameters, created by interpolating the 400~K and 450~K log $g = 4.5$ [m/H] $= -0.5$ grid models (olive lines). The agreement with the observed 
$4.0 \lesssim \lambda~\mu$m $\lesssim 18$
spectrum could be made acceptable with an increase in the scaling factor of $\approx 25\%$, however this would increase the discrepancy in the near-infrared where the model flux is a factor of $\sim 3$ too bright at $J$ and $H$. Improving the agreement in the near-infrared would require a larger deviation from an adiabatic pressure-temperature relationship in order to cool the interior atmosphere.
   \item 
 \citet{Barrado_2023} used ATMO 2020++ models \citep{Leggett_2021, Leggett_2023, Meisner_2023}  with the 
 $5.0 \lesssim \lambda~\mu$m $\lesssim 18$ MIRI data to determine 
$T_{\rm eff} = 400~K$, log $g = 4.0$, and [m/H] $= 0.0$ (green line in Figure 5). The bottom panel of Figure 5 shows that while the agreement with the MIRI data is very good, the flux in the 4.5~$\mu$m region is approximately half what is observed. That is, adoption of a lower gravity increases the CO and CO$_2$ absorption at $4.7~\mu$m and $4.25~\mu$m and
produces a spectrum that does not reproduce the NIRSpec data; the W2 $-$ W3 color is too red, as also demonstrated in Figure 4.
Furthermore, the discrepancy in the near-infrared is around a factor of 5 (Figure 5).
Barrado et al. also use the ARCiS framework \citep{Ormel_2019} for a self-consistent atmospheric analysis of the MIRI data, and determine  $T_{\rm eff} = 500~K$ and log $g = 5.0$. Figure 4 indicates that a brown dwarf this warm will be more than ten times brighter than observed, at near-infrared wavelengths.
\end{itemize}
\citet{Lew_2024} and \citet{Barrado_2023} also explore retrieval analyses. The Lew et al. CHIMERA \citep{Line_2014} retrieval result of $T_{\rm eff} = 530~K$ and log $g = 5.2$, like the Barrado et al. ARCiS self-consistent result, will be incompatible with the observed near-infrared flux (Figure 4).
\citet{Barrado_2023} use ARCiS \citep{Ormel_2019}, Brewster \citep{Burningham_2017}, and petitRADTRANS \citep{Molliere_2019} for their retrieval analyses of the MIRI data. They find solar metallicity solutions with
$T_{\rm eff}$ values between 350~K and 400~K, and log $g$ values between 3.8 and 4.7; they also find a metal-poor solution with a very low gravity of 3.4 dex.  The ATMO 2020++ models indicate that the lower gravity solutions will be inconsistent with the NIRSpec data, as shown spectroscopically in the bottom panel of Figure 5, and photometrically in Figure 4. The higher gravity solutions are consistent with our results, and the \citet{Barrado_2023} radii are consistent with evolutionary models if the source is an unresolved binary.

In summary, broad wavelength coverage is required for reliable analyses of brown dwarf spectra. Taken separately, the near-infrared region, the 4.5~$\mu$m region, and the 12~$\mu$m region,  can each indicate atmospheric properties that are excluded by data at other wavelengths. While our preferred ATMO 2020++ solutions show some discrepancies at $\lambda \approx 1.6~\mu$m and   $\lambda \approx 4.0~\mu$m (Figure 5, top two panels), they provide an excellent fit across the spectral regions where significant flux is emitted (Figure 6).

Adopting a $T_{\rm eff} = 400$~K and log $g = 5.0$ binary solution,  evolutionary models  then imply an age of 8 to 15~Gyr, corresponding to masses of 20 to 30~$M_{\rm Jup}$ respectively, for each component \citep{Marley_2021}. The tangential velocity of 48~km s$^{-1}$ suggests thin disk membership \citep[Figure 31 of][]{Dupuy_2012}, and so an age at the younger end of this range, and a mass at the smaller end, is more likely. This is a tight system -- \citet{Furio_2023} find that the separation must be smaller than 0.5~au, or around 1000 brown dwarf radii. An order-of-magnitude estimate by \citet{Lew_2024}, based on the radial velocity, suggests an even smaller separation of 20 Jupiter radii.

This enigmatic source therefore seems to be revealing itself as a high gravity, massive,  tightly bound, brown dwarf binary. The analysis here also supports the use of the W2 $-$ W3 color (or similar passbands) as a gravity indicator.

\bigskip
\section{Application to {\it JWST} Imaging}

Figure 7 shows a color-magnitude diagram using {\it JWST} filters, similar to the top panel of Figure 4. Spectral coverage of each of these wide bandpasses is illustrated in Figure 2. The F1130W filter was not considered as the bandpass is narrower and the sensitivity is reduced. 

%\clearpage
\begin{figure}[!hb]
\vskip -0.4in
\hskip 0.4in
\includegraphics[angle=0,width = 5.5 in]
{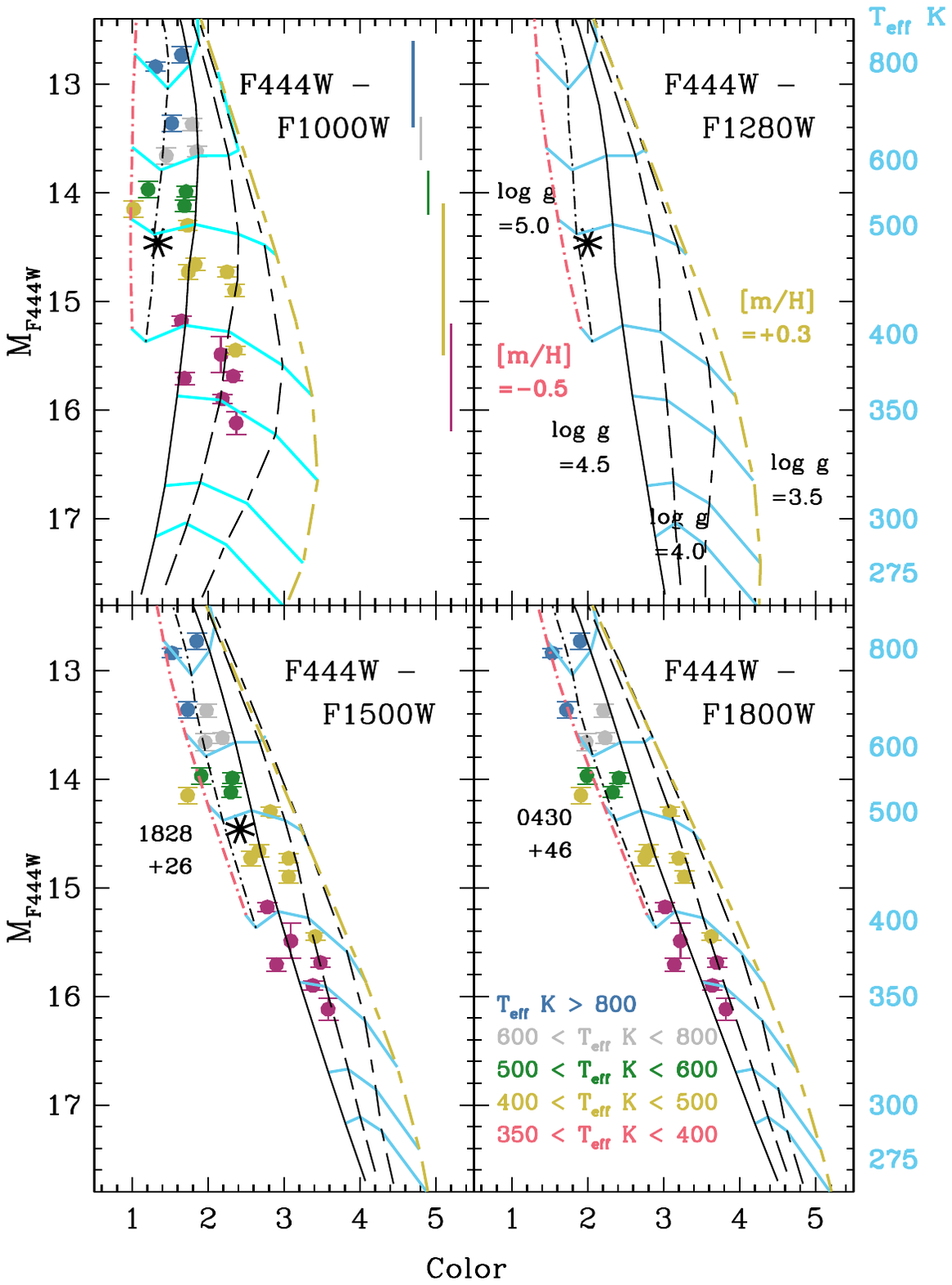}
\vskip -0.2in
\caption{Color-magnitude diagram for late T and Y dwarfs, using {\it JWST} filters.   Filled circles represent F444W, F1000W, F1500W and F1800W colors from  \citet{Beiler_2024}. The symbol color indicates the luminosity-based $T_{\rm eff}$ value determined by Beiler et al., see the legend in the lower right. Colored bars in the upper left illustrate the range in $M_{\rm F444W}$ for each temperature group. The black asterisk represents WISE 1828, using F444W, F1000W, F1280W and F1500W colors synthesized here from {\it JWST} spectra. Vertical lines are ATMO 2020++ PH$_3$-free sequences, with cyan lines showing isotherms for the model $T_{\rm eff}$ on the right axes.  
The black lines are log $g =$ 3.5, 4.0, 4.5, and 5.0 from right to left. The red line is a log $g =$ 5.0 sequence with [m/H] $= -0.5$ and the yellow line is a log $g =$ 3.5 sequence with [m/H] $= +0.3$.  
The source with the bluest F444W $-$ F1000W color is  WISE J043052.92+463331. 
}
\end{figure}

We synthesized F444W, F1000W, F1280W, and F1500W magnitudes for WISE 1828 using the NIRSpec and MIRI spectra published by \citet{Lew_2024} and \citet{Barrado_2023}. The MIRI spectrum does not cover the entire F1800W bandpass. We determined: F444W $= 14.45$, F1000W $= 13.11$, F1280W $= 12.45$, and F1500W $= 12.03$. The uncertainty is dominated by the absolute flux uncertainty in the spectra, currently estimated to be around 5\% \footnote{\url{https://jwst-docs.stsci.edu/jwst-calibration-status/jwst-absolute-flux-calibration\#gsc.tab=0
}}.

Also shown in Figure 7 are F444W, F1000W, F1500W, and F1800W colors for 22 T and Y dwarfs from \citet[][their Tables 1 and 9]{Beiler_2024}. The F444W magnitudes are synthesized from NIRSpec spectra, while F1000W, F1500W, and F1800W are from MIRI imaging observations.   The Beiler et al. sample of brown dwarfs appears to be typical of the local disk, with the T dwarfs ($T_{\rm eff} \gtrsim 500$~K) having log $g$ values of 4.5 to 5.0, and the Y dwarfs ($T_{\rm eff} \lesssim 500$~K) having log $g$ values of 4.0 to 4.5, both corresponding to ages of a few Gyr \citep[][their Figure 10]{Marley_2021}.
While there are no extremely low-gravity sources in 
the sample, one source is likely to be high gravity and metal-poor: WISE J043052.92+463331. This T8 dwarf has the bluest F444W $-$ F1000W color in the \citet{Beiler_2024} sample and has been flagged as abnormally red in $H - $W2, and faint in $H$ by \citet{Best_2021}. 

The models are in excellent agreement with the observations --- both in terms of the colors and the $T_{\rm eff}$ values, which were determined by  \citet{Beiler_2024} using the 
observed bolometric luminosity. Figure 7 shows that the shorter wavelength colors -- F444W $-$ F1000W and F444W $-$ F1280 -- are more sensitive to gravity than the longer wavelength colors, however
F444W $-$ F1500W and F444W $-$ F1800W continue to increase with decreasing $T_{\rm eff}$, making them useful for temperature estimates in the absence of a distance measurement.  
To ease comparisons using different $4.5~\mu$m filter bandpasses, 
\citet[][their Figure 6 and Table 3]{Leggett_2023} provide transformations determined from synthetic spectra; the differences are relatively small -- around 0.2 magnitudes for late-T and Y dwarfs.

\bigskip
\section{Conclusions}

Separating metallicity and surface gravity effects in near-infrared spectra of cool brown dwarfs has been a long-standing challenge \citep[e.g.][]{Burgasser_2006b,Liu_2007,Leggett_2017}. Inspired by clear trends seen at $4 \lesssim \lambda~\mu$m $\lesssim 15$ in early fits of {\it JWST} Y dwarf spectra \citep{Leggett_2024}, we explored the sensitivity of late-T and Y dwarf {\it WISE} W2 and W3 colors to $T_{\rm eff}$, [m/H], and log $g$, after reviewing and updating the W3 magnitudes of 82 objects (Section 2).  We use PH$_3$-free ATMO 2020++ models which have been shown to reproduce {\it JWST} spectroscopic observations of late-T and Y dwarfs over a broad wavelength range of $1 \lesssim \lambda~\mu$m $\lesssim 14$ \citep{Leggett_2023,Leggett_2024, Tu_2024}. The  $T_{\rm eff}$ values
determined using ATMO 2020++ synthetic spectra are in good agreement with values determined semi-empirically by \citet{Beiler_2023}  using bolometric luminosity arguments (Figure 7). Furthermore, the  $T_{\rm eff}$ and log $g$ values are consistent with evolutionary models \citep[e.g.][]{Marley_2021}, assuming this local sample of brown dwarfs has an age range of around 0.5~Gyr to 6~Gyr \citep{Dupuy_2017, Kirkpatrick_2019, Best_2024}.

We find that for  $T_{\rm eff} \lesssim 600$~K the W2 $-$ W3 color becomes very sensitive to gravity, while being only mildly sensitive to metallicity (Figures 4 and 7). Other observations and analyses support this conclusion:
\begin{itemize}
\item 
We identify ten very red, likely low-gravity, low-mass young brown dwarfs or exoplanets, listed in Table 3. Two of these are companions to main sequence stars and previous analyses \citep{Zhang_2021a,Rothermich_2024} have found the systems to be young (100 -- 800~Myr). Previous near-infrared spectral analysis of a third source supports the low gravity, mass, and age (log $g \approx 3.7$, \citet{Zhang_2021b,Zalesky_2022}). Two additional sources have kinematics indicative of a young age. Followup observations are required to confirm the young and low-mass nature of these objects: CWISE J004143.77$-$401929.9, CWISEP J010650.61+225159.1, CWISE J104053.42$-$355029.7, ULAS J115229.67+035927.2, and WISE J223204.50-573010.5 (Section 3.3). 
\item 
The sources with the bluest W2 $-$ W3 colors, which the models show to be indicative of high gravity and metal paucity, are known metal-poor T dwarfs, and are likely to be around 8~Gyr old with masses around 35~$M_{\rm Jup}$ (Section 3.4).
\item 
The revised W3 magnitude for the Y dwarf WISEPA J182831.08+265037.8 places it along the high gravity W2 $-$ W3 sequence in Figure 4. We find excellent agreement with the observed 4 to 18~$\mu$m {\it JWST} spectra \citep{Barrado_2023, Lew_2024} if the system is a pair of similar cold brown dwarfs with $T_{\rm eff}$ 350~K to 400~K, surface gravities $g$ in the range log $g$ 4.5 to 5.0 dex, and a solar or slightly sub-solar metallicity. Evolutionary models then imply an age around 8~Gyr and a mass $\sim$20 $M_{\rm Jup}$ for each component.
The separation is less than 1000 radii based on unresolved {\it JWST} images \citep{Furio_2023}.
\end{itemize}
Other evidence therefore supports the model trends shown in Figure 4, illustrating the power of a color which measures the slope of the energy distribution from $\lambda \sim 5~\mu$m to $\lambda \sim 10~\mu$m.

The {\it WISE} W3 magnitude is difficult to measure to better than around 20\% for sources with W3 $\approx 13$, due to the large pixels of the camera and the high and spatially-variable background. MIRI imaging with {\it JWST} offers the opportunity  to reach a higher precision and fainter sources. 
\citet{Beiler_2024} determine accurate mid-infrared colors  ($\pm ~3\%$) and $T_{\rm eff}$ values ($\pm$ 4\% to 7\%) for a sample of 22 T and Y brown dwarfs as faint as 14th magnitude at $\lambda = 10~\mu$m.
Figure 7 shows that the ATMO 2020++ models are
in excellent agreement with the data, 
illustrating the potential of combining {\it JWST} data with these models.  For example, the models demonstrate that the Beiler et al. sample has a range in surface gravity which corresponds to a local disk sample aged 0.5~Gyr to 6~Gyr \citep[using the evolutionary models of][]{Marley_2021}, and a     
small range in  metallicity, typical of the local solar neighborhood \citep[e.g.][]{Hinkel_2014}.

Future mid-infrared missions with sufficient sensitivity at $4 \lesssim \lambda~\mu$m $\lesssim 15$ will be able to identify old or very metal-poor brown dwarfs in the solar neighborhood, as well as exoplanets and planetary-mass free-floating brown dwarfs. The Near-Earth Object Surveyor Mission, scheduled for launch in 2027 or 
2028\footnote{\url{https://neos.arizona.edu/}} is designed to detect 200~K to 300~K objects in filters centered near $4.6~\mu$m and $8~\mu$m \citep{Mainzer_2023}. The $8~\mu$m sensitivity of $\sim 5$~mJy, illustrated in \citet{Mainzer_2023}, suggests that a detection in both filters is limited to young low-mass 350~K -- 400~K brown dwarfs which  are closer than $\sim 4$~pc, based on the ATMO 2020++ models. Current space density estimates of Y dwarfs \citep{Kirkpatrick_2021a, Best_2024} then suggest that zero to 1 dwarf will be detected. However these estimates are very uncertain and the mission remains of great interest for brown dwarf and exoplanet science, as well as solar system science.

The Appendix provides the photometric data compilation used in this work, which is described in Section 2. The model spectra and colors are publicly available at 
https://opendata.erc-atmo.eu, under the heading ``ATMO 2020++ grid without PH3, Leggett \& Tremblin (2024)''.

\begin{acknowledgements}

We are very grateful to the referee, whose comments significantly improved this paper.

We are beyond grateful to the many engineers and scientists who made the NIRSpec and MIRI instruments, and the {\it JWST}, the success that they are. 

This research has made use of the NASA/IPAC Infrared Science Archive, which is funded by the National Aeronautics and Space Administration and operated by the California Institute of Technology.

\end{acknowledgements}

\appendix

%\clearpage
\bigskip
\section{Photometry Compilation}

Table 4 presents a compilation of photometry.

\clearpage
\setlength\tabcolsep{1pt}
%\movetabledown=2.8in
\movetabledown=2.5in
\begin{rotatetable}
\begin{deluxetable*}{lcccrrrrrrrrrrrrrrrrrrrrrrccccc}
\tabletypesize{\tiny}
\tablecaption{Compilation of Photometric Measurements for T- and Y-Type Brown Dwarfs}
\tablehead{
\\
\colhead{Survey} & 
\multicolumn{2}{c}{Discovery RA Decl.}  & 
%\colhead{Other} & 
\colhead{Sp.} & 
\colhead{$M-m$} & 
\colhead{$Y$} & \colhead{$J$} & \colhead{$H$} & \colhead{$K$} & \colhead{$L^{\prime}$} &
% \colhead{$M^{\prime}$} & 
%\colhead{IR Note} &
\colhead{[3.6]} & \colhead{[4.5]} &
% \colhead{[5.8]} & \colhead{[8.0]} & 
\colhead{W1} & \colhead{W2} & \colhead{W3} & 
%\colhead{W4} & 
%\colhead{WISE Note} &
\colhead{$e_{Mm}$} & 
\colhead{$e_Y$} & \colhead{$e_J$} & \colhead{$e_H$} & \colhead{$e_K$} & \colhead{$e_{L^{\prime}}$} & %\colhead{$e_{M^{\prime}}$} & 
\colhead{$e_{3.6}$} & \colhead{$e_{4.5}$} & 
%\colhead{$e_{[5.8]}$} & \colhead{$e_{[8.0]}$} &
\colhead{$e_{W1}$} & \colhead{$e_{W2}$} & \colhead{$e_{W3}$} & 
%\colhead{$e_{W4}$} & 
\multicolumn{5}{c}{References}\\
\colhead{Name} &
\colhead{hhmmss.ss} & \colhead{$\pm$ddmmss.s} & 
%\colhead{Names} & 
\colhead{Type} & 
\multicolumn{22}{c}{mag} & 
\colhead{Discovery} & \colhead{Sp. Type} & \colhead{Parallax} & \colhead{Near-IR} & \colhead{$Spitzer$} 
}
\startdata
CWISEP &   000229.93 & +635217.0     &  7.5    &      &         &      &        &       &      &  17.35  & 15.69 &               18.21  &   15.73     &      &   &   &   &   &   &   & 0.25   & 0.06     &  0.21   &   0.05    &  &                                            Meisner${\_}$2020b  &  Meisner${\_}$2020b &  & & Meisner${\_}$2020b  \\            
WISE   & 000517.48 & +373720.5       &   9.0  & 0.52  &  18.48  & 17.59 & 17.98 & 17.99 & 14.43 & 15.43  & 13.28 & 16.51 & 13.30 & 11.75 &
                                      0.04  & 0.02  & 0.02  & 0.03  & 0.03  & 0.10  & 0.04  & 0.04   &  0.04  & 0.01  & 0.08   &      
    Mace${\_}$2013a   &    Mace${\_}$2013a     &      Kirkpatrick${\_}$2019    &        Leggett${\_}$2015    &           Kirkpatrick${\_}$2019 \\               
CWISEP   &    001146.07 & -471306.8   &    8.5  &   &    & 19.28  &  19.69    &     &         &    17.74  & 15.81  &  18.75 & 15.99 &   &
                               &  &        0.07 & 0.20   & & &  0.09  &  0.02   & 0.20 &  0.05  &      &        
    Meisner${\_}$2020a                 &            Meisner${\_}$2020a        &    &   VISTA${\_}$VHS     &                Meisner${\_}$2020a   \\                
WISE   &      001354.39 & +063448.2   &   8.0   &    & 20.56  & 19.54  & 19.98  & 20.79    &      &    17.15 &  15.16  & 17.95 &   15.23 & &
                                             & 0.04 & 0.03 & 0.04 & 0.10 &   &           0.03  &   0.03   & 0.11 &       0.03 & &
    Pinfield${\_}$2014a            &                 Pinfield${\_}$2014a                   &                     &             Leggett${\_}$2015       &   Pinfield${\_}$2014a    \\              
WISEA   &     001449.96 & +795116.1    &   8.0  &     &   20.32  & 19.36   &       &     &       &     17.76 &  15.88 &  19.32 & 15.97 &  13.55  &     
                     &   0.10 & 0.10 &   &   &   &   0.04  &    0.02 &    0.28  & 0.04 &  0.27 &
                     Bardalez${\_}$2020        &                    Bardalez${\_}$2020             &    &     Leggett${\_}$2021      &        Bardalez${\_}$2020                   \\
\enddata
\tablecomments{Table 4 is published in its entirety in the machine-readable format.
      A portion is shown here for guidance regarding its form and content.}
\vskip -0.05in
\tablerefs{
2MASS -- \citet{Skrutskie_2006};
Albert${\_}$2011 -- \citet{Albert_2011};
Artigau${\_}$2010 -- \citet{Artigau_2010};
Bardalez${\_}$2020 -- \citet{Bardalez_2020};
Beichman${\_}$2014 -- \citet{Beichman_2014}; 
Beiler${\_}$2024 -- \citet{Beiler_2024};
Best${\_}$2015 -- \citet{Best_2015};
Best${\_}$2020 -- \citet{Best_2020};
Best${\_}$2021 -- \citet{Best_2021}; 
Brooks${\_}$2022 -- \citet{Brooks_2022};
Burgasser${\_}$1999 -- \citet{Burgasser_1999};
Burgasser${\_}$2000 -- \citet{Burgasser_2000};
Burgasser${\_}$2002-- \citet{Burgasser_2002};
Burgasser${\_}$2003 -- \citet{Burgasser_2003};
Burgasser${\_}$2004-- \citet{Burgasser_2004};
Burgasser${\_}$2006 -- \citet{Burgasser_2006a};
Burgasser${\_}$2008-- \citet{Burgasser_2008};
Burgasser${\_}$2010-- \citet{Burgasser_2010}
Burgasser${\_}$2012-- \citet{Burgasser_2012}
Burningham${\_}$2008 -- \citet{Burningham_2008};
Burningham${\_}$2009 -- \citet{Burningham_2009};
Burningham${\_}$2010a -- \citet{Burningham_2010a};
Burningham${\_}$2010b -- \citet{Burningham_2010b};
Burningham${\_}$2011 -- \citet{Burningham_2011};
Burningham${\_}2$013 -- \citet{Burningham_2013};
Chiu${\_}$2006 -- \cite{Chiu_2006};
Cushing${\_}$2011 -- \citet{Cushing_2011};
Cushing${\_}$2014 -- \citet{Cushing_2014};
Cushing${\_}$2016 -- \citet{Cushing_2016};
Delorme${\_}$2008 -- \citet{Delorme_2008};
Delorme${\_}$2010 -- \citet{Delorme_2010};
Dupuy${\_}$2012 -- \citet{Dupuy_2012};
Dupuy${\_}$2012 -- \citet{Dupuy_2013};
Dupuy${\_}$2015 -- \citet{Dupuy_2015};
Faherty${\_}$2012 -- \citet{Faherty_2012};
Faherty${\_}$2020 -- \citet{Faherty_2020};
Gaia -- \citet{GAIA};
Geballe${\_}$2001  -- \citet{Geballe_2001};
Gelino${\_}$2011 -- \citet{Gelino_2011};
Goldman${\_}$2010 -- \citet{Goldman_2010};
Golimowski${\_}$2004 -- \citet{Golimowski_2004};
Greco${\_}$2019 -- \citet{Greco_2019};
Griffith${\_}$2012 -- \citet{Griffith_2012};
Kirkpatrick${\_}$2011 -- \citet{Kirkpatrick_2011};
Kirkpatrick${\_}$2012 -- \citet{Kirkpatrick_2012};
Kirkpatrick${\_}$2013  -- \citet{Kirkpatrick_2013};
Kirkpatrick${\_}$2019  -- \citet{Kirkpatrick_2019};
%Kirkpatrick{\_}2020  -- \citet{Kirkpatrick_2020};
Kirkpatrick${\_}$2021a  -- \citet{Kirkpatrick_2021a};
Kirkpatrick${\_}$2021b  -- \citet{Kirkpatrick_2021b};
Knapp${\_}$2004 -- \citet{Knapp_2004};
Kota${\_}$2022 -- \citet{Kota_2022};
Leggett${\_}$2002 - \citet{Leggett_2002};
Leggett${\_}$2009 - \citet{Leggett_2009};
Leggett${\_}$2010 - \citet{Leggett_2010};
Leggett${\_}$2012 - \citet{Leggett_2012};
Leggett${\_}$2013 - \citet{Leggett_2013};
Leggett${\_}$2014 - \citet{Leggett_2014};
Leggett${\_}$2015 - \citet{Leggett_2015};
Leggett${\_}$2017 - \citet{Leggett_2017};
Leggett${\_}$2019 - \citet{Leggett_2019};
Leggett${\_}$2021 - \citet{Leggett_2021};
Liu${\_}$2011 - \citet{Liu_2011};
Liu${\_}$2012 - \citet{Liu_2012};
Lodieu${\_}$2007 -- \citet{Lodieu_2007};
Lodieu${\_}$2009 -- \citet{Lodieu_2009};
Lodieu${\_}$2012 -- \citet{Lodieu_2012};
Lodieu${\_}$2022 -- \citet{Lodieu_2022};
Looper${\_}$2007 -- \citet{Looper_2007};
Lucas${\_}$2010 -- \citet{Lucas_2010};
Luhman${\_}$2011 -- \citet{Luhman_2011};
Luhman${\_}$2012 -- \citet{Luhman_2012};
Luhman${\_}$2014 -- \citet{Luhman_2014};
Mace${\_}$2013a -- \citet{Mace_2013a}; 
Mace${\_}$2013b -- \citet{Mace_2013b}; 
Mainzer${\_}$2011 -- \citet{Mainzer_2011};
Manjavacas${\_}$2013 -- \citet{Manjavacas_2013};
Marocco${\_}$2010 -- \cite{Marocco_2010};
Marocco${\_}$2020 -- \cite{Marocco_2020};
Martin${\_}$2018 -- \citet{Martin_2018};
Meisner${\_}$2020a -- \citet{Meisner_2020a};
Meisner${\_}$2020b -- \citet{Meisner_2020b};
Meisner${\_}$2021 -- \citet{Meisner_2021};
Meisner${\_}$2023 -- \citet{Meisner_2023};
Meisner${\_}$2024 -- \citet{Meisner_2024};
Murray${\_}$2011 -- \citet{Murray_2011};
Patten${\_}$2006 -- \citet{Patten_2006};
Pinfield${\_}$Gromadzki${\_}$2014 -- \rm{Pinfield, P. and Gromadzki, M. private communication 2014;}
Pinfield${\_}$2008 -- \citet{Pinfield_2008};
Pinfield${\_}$2012 -- \citet{Pinfield_2012};
Pinfield${\_}$2014a -- \citet{Pinfield_2014a};
Pinfield${\_}$2014b -- \citet{Pinfield_2014b};
Robbins${\_}$2023     -- \citet{Robbins_2023};      Rothermich${\_}$2024    -- \citet{Rothermich_2024};   Schneider${\_}$2015     -- \citet{Schneider_2015};    Schneider${\_}$2020      -- \citet{Schneider_2020};                 Schneider${\_}$2021  -- \citet{Schneider_2021};
Scholz${\_}$2010a -- \citet{Scholz_2010a};
Scholz${\_}$2010b -- \citet{Scholz_2010b};
Scholz${\_}$2011 -- \citet{Scholz_2011};
Scholz${\_}$2012 -- \citet{Scholz_2012};
Smart${\_}$2010 -- \citet{Smart_2010};
Strauss${\_}$1999 -- \cite{Strauss_1999};
Subasavage${\_}$2009 -- \citet{Subasavage_2009};
Thompson${\_}$2013 -- \citet{Thompson_2013};
Tinney${\_}$2003 -- \citet{Tinney_2003};
Tinney${\_}$2005 -- \citet{Tinney_2005};
Tinney${\_}$2012 -- \citet{Tinney_2012};
Tinney${\_}$2014 -- \citet{Tinney_2014};
Tinney${\_}$2018 -- \citet{Tinney_2018};
Tsvetanov${\_}$2000 -- \citet{Tsvetanov_2000};
TW -- This Work;
UKIDSS -- \citet{Lawrence_2007};
VISTA -- \citet{Sutherland_2015};
Vrba${\_}$2004 -- \citet{Vrba_2004};
Warren${\_}$2007 -- \citet{Warren_2007};
Wright${\_}$2013 -- \citet{Wright_2013}
}      
\end{deluxetable*}
\end{rotatetable}

\clearpage
\bibliography{Leggett_Tremblin_Nov2024}{}
\bibliographystyle{aasjournal}

\end{document}